\documentclass{tADP2e}
\usepackage{bm, graphicx}

\def\he4{$^4$He}
\def\hee3{$^3$He}
\def\Am3{\AA$^{-3}$}
\def\beq{\begin{equation}}
\def\eeq{\end{equation}}

\begin{document}

\title{{\itshape What makes a crystal supersolid ?}
}

\author{Nikolay Prokof'ev,
Department of Physics, University of Massachusetts,
Amherst, MA 01003, USA }


\maketitle
\begin{abstract}
For nearly half a century the supersolid  phase of matter has
remained mysterious, not only eluding experimental observation, but
also generating a great deal of controversy among theorists. Recent
discovery of what is interpreted as a non-classical moment of
inertia at low temperature in solid $^4$He \cite{KC1,KC1b,KC2,KC3}
has elicited much excitement as a possible first observation of a
supersolid phase.  In the two years following the discovery,
however,  more puzzles than answers have been provided to the
fundamental issue of whether the supersolid phase exists, in helium
or any other naturally occurring condensed matter system. Presently,
there is no established theoretical framework to understand the body
of experimental data on \he4. Different microscopic mechanisms that
have been suggested to underlie superfluidity in a perfect quantum
crystal do not seem viable for \he4, for which a wealth of
experimental and theoretical evidence points to an insulating
crystalline ground state. This perspective addresses some of the
outstanding problems with the interpretation of recent experimental
observations of the apparent superfluid response in \he4 (seen now
by several groups \cite{Rittner,Shirahama,Kubota}) and discusses
various scenarios alternative to the homogeneous supersolid phase,
such as superfluidity induced by extended  defects of the
crystalline structure which include grain boundaries, dislocations,
anisotropic stresses, etc. Can a metastable superfluid ``glassy"
phase exist, and can it be relevant to some of the experimental
observations ? One of the most interesting and unsolved fundamental
questions is what interatomic potentials, given the freedom to
design one, can support an ideal supersolid phase in continuous
space, and can they be found in Nature.
\end{abstract}

\section{Introduction}
\label{sec:Int}

The textbook notion of a perfect crystal at $T=0$, is that of a periodic array
of unit cells, all comprising the same {\it integer} number of particles,
$\nu = \int_{\Omega }d{\bm r} \rho ({\bm r})$, where $\Omega$
is the unit cell volume and $\rho ({\bm r})$ is the average particle density
profile. For simplicity, let us consider a single component crystal
similar to $^4$He. We assume that the elementary constituents (atoms or molecules)
can be regarded as structureless particles; furthermore, we assume that they obey Bose
statistics. 

\begin{figure}[t]
\centerline{\includegraphics
[angle=-90, scale=0.5]{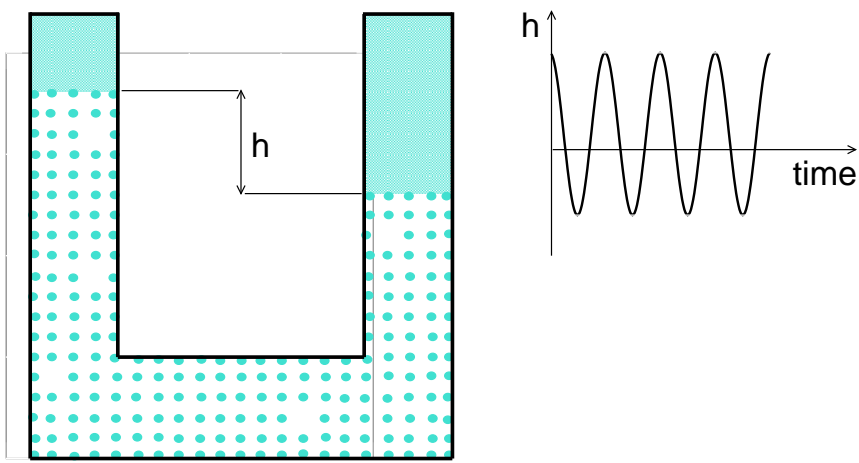}}
\vspace*{-3cm}
\caption{Supersolid crystal (represented by periodic arrays
of dots) in coexistence with the superfluid liquid of lower density
(shaded regions on top of the U-shaped closed tube) in the gravity field.
If initial conditions are such that supersolid
levels to the left and to the right are at different height, the system
will start oscillating by melting on one side and recrystallizing
released density on another side  with coherent dissipationless
mass flow through the otherwise static crystalline structure.
[The assumption is that the flow velocity remains below the critical value,
or, $h$ is small enough,
and the supersolid-liquid interface is rough.]}
\label{fig1}
\end{figure}

The supersolid phase (SFS) can be generally defined as one that combines
crystalline properties, such as shear modulus and broken translation symmetry,
with frictionless mass transport through the solid bulk.
The striking, simultaneous presence of solid and superfluid properties in
the same condensed matter system, will result in a number of phenomena
that defy our  everyday experience. One such phenomenon is
schematically depicted in Fig.~\ref{fig1}, showing a SFS sample
(region with periodic arrays of dots) placed inside a U-shaped vessel,
in coexistence with the superfluid liquid. At $T=0$ undamped oscillations
of the solid levels in the two sections of the U-shape vessel will be observed,
as long as  the maximum flow velocity remains below the critical value
and the liquid-solid interface is rough.

Another landmark of SFS behavior (see, for instance, Ref. \cite{Leggett70}),
is the reduction of the moment of inertia of a solid sample with respect
to its classical value (commonly referred to as Non-Classical Rotational
Inertia, or NCRI) 
\begin{equation}
I(T)=I_{\rm class}(1-n_s(T)/n)\;.
\label{NCRI}
\end{equation}
Here $n_s(T)$ is the superfluid density, and $n$ is the particle 
density in the system. The NCRI effect can be observed by enclosing a 
known amount of solid in a vessel, which is then set in rotation about its axis. 
The moment of inertia of the system is related to the (measurable) resonant 
period of oscillation of the rotating system. A drop of the moment of inertia
at low temperature, is interpreted as {\it decoupling} of part of the solid
(the superfluid fraction) from the rotation.

\begin{figure}[t]
\includegraphics [angle=-90, scale=0.4]{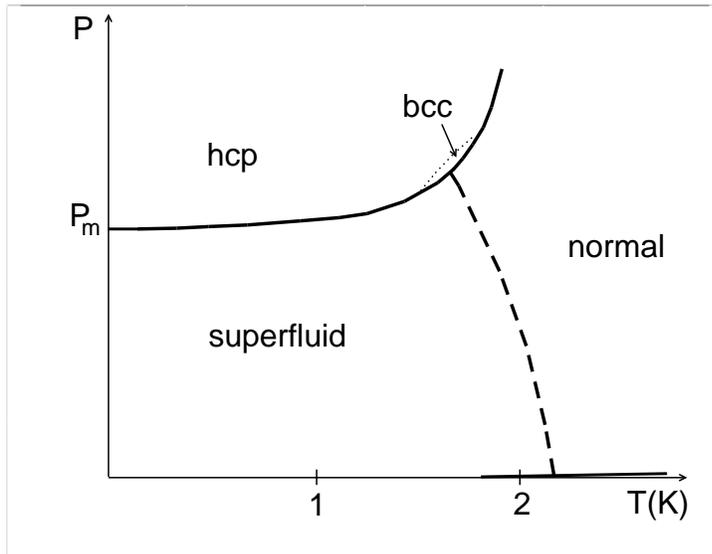}
\vspace*{-1cm}
\caption{A sketch of the \he4 phase diagram. At low temperature
the helium superfluid undergoes a weak first-order phase transition to the
hexagonal close-packed (hcp) solid phase at pressure $P_m \approx 25.6$~bar. }
\label{fig0}
\end{figure}

The history of ideas on how superfluidity can occur in a crystalline solid
is quite old. On very general grounds, one expects that the ``textbook''
crystal ought to be insulating, i.e., with vanishing superfluid density 
$n_s$, and off-diagonal correlations decaying exponentially in space. 
Penrose and Onsager (PO) were the first to argue in favor of this point 
of view \cite{Penrose51,Penrose}, starting from the picture of atoms 
localized around their equilibrium lattice positions. 
However, their argument has never been regarded as the final word
in the discussion of whether the superfluid solid
phase of matter is possible. C. N. Yang made a comment that long-range phase 
correlations in the solid state could occur in systems characterized 
by a high degree of atomic delocalization \cite{Yang}. Helium is unique in this
regard; quantum properties of light helium atoms in combination with
the relatively weak interatomic potential (the well depth is only
$\sim 10$ K) prevent them from making a crystal at zero pressure, 
see Fig.~\ref{fig0}. An objection to the PO treatment was that regarding
atoms as being localized around lattice points underestimates the role
of exchange processes, which may allow for the (perhaps remote) possibility
of superflow, even with an integer  $\nu$ \cite{Leggett70,Leggett04}.

Another proposal for the supersolid phase has its origin in the observation
that there is no fundamental reason why crystals should necessarily 
be {\it commensurate}, i.e., feature an integer number of particles 
per unit cell, on average.  
A dilute gas of vacancies, or interstitials, which occur 
in all solids at finite temperature and ought to be highly mobile in a 
helium crystal, could be also present in the ground state giving rise to
Bose-Einstein condensation and superfluidity at low temperature
\cite{Andreev69,Chester}. 

Though certainly plausible, most theoretical scenarios leading to the 
SFS phase are phenomenological and qualitative when it comes to making
predictions for a particular Hamiltonian; they have never been
supported by robust microscopic calculations, including all quantum-mechanical
effects and based on realistic models of crystals.
\footnote{A major current area of investigation, in the broader context of 
the study of exotic phases of matter, focuses on quantum {\it lattice}
models. Supersolid phases have been predicted to occur in some of these models \cite{yes_SFS1,yes_SFS2} and may soon be observed experimentally, e.g. in optical lattices.
Albeit of clear fundamental interest, this topic is {\it not} part of the
discussion in this paper. We also do not discuss multicomponent cases in which
one component forms a solid and, thus, acts as an external periodic potential
for the other. Such cases are expected to behave similarly to lattice models.}
Still, their appeal is so strong that they inspired decades of active research
and continue to do so. 

Remarkable manifestations of quantum behavior on a macroscopic scale in supersolids
are expected to render the observation of the SFS state rather unambiguous,
as long as it occurs under conditions (e.g., temperature) accessible
to the experiment. Such an observation has eluded experiments for some
thirty years (for a review of the experimental evidence prior to 2004, see,
for instance, Ref. \cite{Meisel}); it was only in 2004 that the first
convincing evidence of a SFS phase of \he4 was published. The phenomenon was
first reported for a solid embedded in a porous Vycor matrix by
\cite{KC1} Kim and Chan (KC);
shortly thereafter, it was extended to the system embedded in a matrix of porous gold
(with a characteristic pore size two orders of magnitude greater than in Vycor),
and eventually to the bulk crystal \cite{KC1b}.
In all cases, the observed NCRI at low temperature was identified as
the onset of supersolidity in \he4. It should be emphasized that the
supersolid interpretation of the data is supported by the 
velocity/amplitude dependence of the NCRI signal and the  
crucial test that the NCRI effect goes away (is suppressed by two orders of magnitude)
when the flow channel around the rotation axis is blocked. These results
constitute a major problem for the non-superfluid kinetic interpretation.

The Kim and Chan discovery sparked a renewed effort in the investigation of the SFS phase
of matter. At the time of this writing, it seems fair to state that things
have turned out to be considerably more complicated (and thus more interesting)
than expected. As the initial experiment by Kim and Chan is being repeated by
other groups, and concurrently different experiments and reliable microscopic
(first-principles) calculations, based on state-of-the-art computational
techniques are being  carried out, consensus is building that the early
microscopic scenarios of supersolidity, described above, are not viable to
explain the body of experimental data. For example, there is now
strong theoretical and experimental evidence that \he4 {\it is} a commensurate
crystal, and such crystals made of a single species of particles obeying
Bose statistics (e.g., \he4 atoms) are always insulating (more precisely,
the commensurate SFS phase has zero probability of being observed).
Measurements and calculations of vacancy and interstitial properties
indicate that the Andreev-Lifshits-Chester scenario of vacancy-induced
supersolidity does not occur in solid helium, even  out of equilibrium.
Therefore, attention is now shifting to a variety of inhomogeneous scenarios,
that may account for many puzzling (and often apparently conflicting)
experimental results.

This paper is organized as follows:  we shall first review the theoretical
framework for supersolidity, including the early suggestions as well as some
more recent proposals of microscopic mechanisms.
They will be discussed in the light of  both early and recent analytical
and numerical results. We shall then examine the core of experimental data,
which at this time includes observations of NCRI as well as flow experiments
\cite{Beamish} and the most recent experiment on grain boundary
superfluidity \cite{Balibar}, and propose that, while a definitive
theoretical explanation of the observed phenomena is not yet available,
there are good reasons to look at disordered and/or inhomogeneous scenarios.
Specifically, we shall illustrate how the effect of disorder can hold the
key to interpreting an important part of the phenomenology.
Finally, we shall outline some possible directions of experimental
and theoretical investigation.

This perspective is a quick response to the fast developing topic,
not a comprehensive review. I apologize if some important contributions
to the field were overlooked. Given
a broad range of opinions and opposing points of view on the subject,
there is certainly a degree of personal bias towards which arguments
and results are more convincing and should be given more ``weight''
in the discussion. The hope is that the story is provocative
enough to stimulate additional research efforts to understand
the remarkable phenomenon of supersolidity.

\section{Supersolid ground state}

Often the discussion starts with the proposal for the supersolid ground state
wave function $\Psi_G$ though it is certainly allowed by the theory that
in a particular material SFS exists only as a finite-temperature phase.
In the Penrose-Onsager picture of atoms localized around equilibrium points
$\{ {\bm R}_i \}$, the ground state wave function is simply a product of localized Wannier
orbitals
\beq
\Psi^{(PO)}_G=\sqrt{\frac{1}{N!}}\; \sum_P \prod_{j=1}^{N}
\varphi({\bm R}_j-{\bm r}_{Pj})
\label{PO}
\eeq
where the sum $\sum_P$ is over all permutations $P$ of particle labels.
Using the definition of the single-particle density matrix at zero-temperature
\beq
n({\bm r},{\bm r}')=\int \dots \int d{\bm r}_2 d{\bm r}_N
\Psi_G({\bm r},{\bm r}_2, \dots ,{\bm r}_N)
\:\Psi_G({\bm r}',{\bm r}_2, \dots ,{\bm r}_N) \;,
\label{density_matrix}
\eeq
and simple properties of the variational state $\Psi^{(PO)}_G$ one readily
obtains (by substituting $\Psi^{(PO)}_G$ for $\Psi_G$ above)
\beq
n({\bm r},{\bm r}')=\frac{1}{N} \sum_{i}
\varphi ({\bm R}_i-{\bm r}) \varphi({\bm R}_i-{\bm r}') \;.
\label{POdm}
\eeq
Since in the $|{\bm r}-{\bm r}'|\to \infty $ limit either ${\bm r}$
or ${\bm r}'$ has to be far from the equilibrium lattice point ${\bm R}_i$, the density
matrix decays to zero exponentially fast for localized Wannier orbitals.
This immediately implies no off-diagonal long-range order (ODLRO) and
thus no superfluidity in the system.

Here, we adopt the following definition of ODLRO: we say that it is present
in the system, if the integral 
\begin{equation}
\int n({\bm r},{\bm r}') d {\bm r} \to \infty \;, 
{\rm ~~~~~~~~~(Def.~~~ODLRO)}
\label{ODLRO}
\end{equation}
diverges in the thermodynamic limit. This definition is {\it different} and 
more general than that based on finite condensate density ($n_0$), 
which requires that the integral be proportional to the system volume, 
in the thermodynamic limit \cite{Penrose,Kohn,Leggett70,Sherrington70}. 
It makes perfect sense to define ODLRO in such a way that superfluidity 
automatically leads to ODLRO and vice versa. In particular, Eq.~(\ref{ODLRO}) allows 
one to talk about ODLRO in two dimensional (2D) systems at finite temperature
(often referred to as topological order \cite{BKT}) for which
the condensate density is zero. 

In this article, we restrict ourselves to the single-particle superfluidity. 
In general, one has to look at the many-body density matrix to see if
there is ODLRO in some finite-order channel. For example, it is certainly 
possible that M atoms form M-molecules and superfluidity happens than at the 
molecular level---in this case one has to study the M-particle density matrix.
Or, molecular hydrogen would be a good example here, by neglecting 
the internal structure of molecules and treating them as point particles
(probably at the expense of introducing complex effective many-body interactions) 
ODLRO can again be discussed in the single-particle ``molecular" density matrix.
There is no fundamental difference between the two pictures. 
 
The superfluid density (formally a tensor) is a linear response coefficient 
which controls the free-energy increase in a system with periodic boundary 
conditions (BC) when these BC are twisted, i.e. 
if $L_{\alpha}$ is the linear system size in direction 
$\alpha = \hat{x},\hat{y},\hat{z}$ we require that 
$\Psi (\dots , r_{\alpha}+L_{\alpha}, \dots ) =
e^{i\varphi_{\alpha}}\:\Psi (\dots , r_{\alpha}, \dots )$  
(there are other definitions of $n_s$, e.g. through the NCRI, Eq.~(\ref{NCRI}),
which can be shown to be identical to the one given below)
\beq
F({\bm \varphi})-F(0)= \sum_{\alpha \alpha'}\frac{(n_s)_{\alpha \alpha'}}{2m}
\frac{V}{L_{\alpha}L_{\alpha'}} \varphi_{\alpha} \varphi_{\alpha'}  \;.
\label{nsdef}
\eeq
Here $V=\prod_{\alpha} L_{\alpha}$ is the system volume. Twisted BC 
can be avoided by introducing instead a gauge phase with the 
gradient $\nabla \phi$ and writing the free-energy density 
(on the largest scales) as
\[
f({\bm \phi})-f(0) = \sum_{\alpha \alpha'}\frac{(n_s)_{\alpha \alpha'}}{2m} 
(\nabla \phi )_{\alpha} (\nabla \phi )_{\alpha'}\;.
\]
Using ``connectivity" arguments \cite{Kohn,Leggett70,Sherrington70}
one can show that the system is sensitive to twisted BC only if it 
has ODLRO defined in Eq.~(\ref{ODLRO}). 

The wave function (\ref{PO}) takes into account zero-point motion
and local exchange processes between identical particles (these effects are
included in the shape of the Wannier orbitals); nevertheless, it describes an insulating
state, as long as $\varphi ({\bm R}_i-{\bm r})$ are localized. By moving away from two
major assumptions made in the construction of the insulating ground state (\ref{PO}),
one arrives at two, apparently quite different possibilities for the supersolid
wave function. If the role of exchange is underestimated in localizing particles
around different lattice points, then a better variational state might
be of the BEC form
\beq
\Psi^{(BEC)}_G=\prod_{i=1}^{N} \left( \frac{1}{\sqrt{N}} \sum _{j=1}^{N}
\varphi({\bm R}_j-{\bm r}_i) \right) \;,
\label{BEC}
\eeq
with macroscopic condensate in the single-particle state
$ \frac{1}{\sqrt{N}}\sum _{j=1}^{N} \varphi({\bm R}_j-{\bm r})$.
The BEC state (as well as $\Psi^{(PO)}_G$) can be further improved
by multiplying it by a many-body correlation factor of the Jastrow form, namely
$\exp \{-\sum_{i<j=1}^{N} u({\bm r}_i-{\bm r}_j) \}$, whose purpose is to account for the strong
repulsive core of the interatomic potential, and suppress the condensate
fraction ($n_0$) to a value much smaller than unity
($n_0$ is already below 10\% in the superfluid \he4 at zero
pressure \cite{Glyde,ceperley95,moroni04}). By construction, the Jastrow
factor ought not alter, nor eliminate the underlying lattice structure.

Equation (\ref{BEC}) offers an alternative view of the supersolid
state, namely that of a superfluid phase with a density wave modulation.
However, it suffers from a  fundamental shortcoming, in that the number of lattice points
$N_L$ is arbitrarily assumed to be the same as the number of particles.
In a superfluid the density-wave parameters of any variational {\it ansatz}
are independent thermodynamic variables, to be selected through energy minimization.
At this point, we recall that superfluidity and ODLRO imply that
part of the system matter is characterized by a classical field component,
for which the notion of particle number can be safely ignored without loss of
generality. [This component represents low-momentum states with large
occupation numbers.] Thus, the condition $N_L=N\times integer$ can be
satisfied only by accident, which is to say that
a commensurate supersolid in continuous space has a probability of zero measure
to be found in Nature. As a result, Eq. (\ref{BEC}) can not describe the
generic ground state of a realistic system \cite{theorem}.
In order to illustrate more vividly how ODLRO relates to the presence 
of gapless vacancies and/or interstitials and incommensurability of a 
single component solid in continuous space, consider the simple example
of a large-amplitude (classical)
standing wave of the electro-magnetic field in the typical optical
table experiment. In this case, the question would never be asked
whether the number of photons per wave period is an integer.

Another possible modification of Eq.~(\ref{PO}), one that retains the
picture of localized orbitals,
consists of allowing the number of particles $N$ to be less than the number of lattice sites $N_L$, i.e., assuming that the ground state has no energy gap
for the creation of vacancies\footnote{The same argument can be used for
interstitials, although typically their energy cost is higher than for vacancies;
henceforth, we shall implicitly make this assumption consistently throughout the paper}
\beq
\Psi^{(AL)}_G=\sqrt{\frac{(N_L-N)!}{N_L!}}\sum_{\{ k_1 \dots k_N \} }
\; \prod_{j=1}^{N} \varphi({\bm R}_{k_j}-{\bm r}_j)\;,
\label{AL}
\eeq
where the the sum is over all possible sets of $N$ lattice points
out of $N_L$ available. Now, the outcome of the calculation for
the density matrix is quite different because in the sum over $i,j$
\beq
n({\bm r},{\bm r}')=\frac{N_L-N}{N_L^2} \sum_{i,j=1}^{N_L}
\varphi ({\bm R}_i-{\bm r}) \varphi({\bm R}_j-{\bm r}') \;.
\label{ALdm}
\eeq
one can always find terms with ${\bm R}_i$ and ${\bm R}_j$ close,
respectively, to
${\bm r}$ and ${\bm r}'$, no matter how large their
separation. If, in the spirit of a tight-binding approach, we move from a description
in terms of continuous space variable
${\bm r}$ and ${\bm r}'$, to that of lattice site positions for the particle,
by introducing site creation (and, analogously, annihilation) operators
$b_l^{\dag}= \int d{\bm r} \ \varphi ({\bm R}_l-{\bm r})\ b_{\bm r}^{\dag}$,
then Eq.~(\ref{ALdm}) can be recast in the following form (taking lattice translation
invariance into account)
\beq
n(l)=\sum_{k} n(k,k+l) =
\sum_{k} \langle b_k^{\dag} b_{k+l}^{\;} \rangle =\frac{N_L-N}{N_L} \equiv n_v   \;.
\label{ALdmv}
\eeq
The final result is consistent with the picture of BEC of a non-interacting
vacancy gas, with dimensionless lattice concentration $n_v$,
introduced by Andreev and Lifshitz \cite{Andreev69}. One has to assume next
that the effective interaction between gapless vacancies is
repulsive, in order to ensure system stability (i.e., to prevent the vacancy gas from collapsing).

As was noted by Reatto \cite{Reatto} and Chester \cite{Chester},
it is not at all necessary to break translational invariance explicitly in the SFS
state with vacancies, by specifying the set of equilibrium
particle positions $\{ {\bm R}_i \}$. An analogy between the Jastrow
wave function
\beq
\Psi^{(J)}_G \propto \exp  \bigg\{-\sum_{i<j=1}^{N} u({\bm r}_i-{\bm r}_j) \bigg\}   \;,
\label{RC}
\eeq
and the partition function of a classical system of interacting particles with pairwise potential
$v^{(cl)}(r)=Tu(r)$ at finite temperature, suggests that among
(\ref{RC}) there are variational states which feature spontaneous
crystalline order, as classical systems are known to crystallize
at sufficiently low temperature. States (\ref{RC}) are also superfluid
since they feature ODLRO, and incommensurate \cite{Reatto}.
Regarding the last property, one may observe that, due to lack of vacancy-interstitial
symmetry in continuous solids, the probability of having $N=N_L$ is of
zero measure \cite{theorem} (an obvious statement for a classical solid at
finite $T$).

Based on all of the above considerations, one might be led to thinking
that the Jastrow wave function may hold the key to the microscopic understanding
of supersolidity. However, there are fundamental problems
associated with Eq.~(\ref{RC}), chiefly that it can not possibly describe
realistic solids of systems with short-range interactions (such as helium),
for which the condensate fraction is tiny (less than 10$^{-3}$).
The reasoning supporting this criticism is offered in the next Section.

The most important conclusion of this Section is that only zero-point
vacancies or interstitials can make an ideal crystal supersolid. There is
no other ideal-crystal scenario for the SFS state though microscopic mechanisms
leading to the incommensurate crystalline groundstate might be very involved.
However, the story does not end here, since we did not question yet
what happens when quantum crystals are not perfect, i.e. contain defects such as
dislocations and grain boundaries, or even loose their crystalline order
completely and form an amorphous solid. It appears, that topological
defects and glasses offer an alternative approach to supersolid phenomenon.
In what follows we examine which of the two possibilities
is most likely to operate in helium.

\section{Insulating {\it hcp} crystals of \he4.}

Existing theories can not answer the question at what pressures and
temperatures one has to look for the SFS phase, or even whether the SFS
phase exists at all in a given Hamiltonian, e.g. in helium.
Several aspects have emerged as necessary conditions that the SFS state has
to satisfy, in order to exist; consequently, any of them can be used to
probe {\it hcp} crystals of helium theoretically. [At the time of this writing,
there is no experimental evidence of structural transitions in solid
helium at low temperatures $T<1K$ and moderate pressures, say $P<200$~bar.]
By definition, the SFS state has a finite superfluid density $n_s$.
Since superfluidity and ODLRO (as defined above) must be present simultaneously,
one can also look at asymptotic
properties of the single-particle density matrix, which is nothing by the
zero time limit $n({\bm r})=G({\bm r},-0)$ of the Matsubara Green function
\begin{equation}
G({\bm r},\tau ) = V^{-1}\int d{\bm r}'  \langle \, {\cal T} \{
\hat b^{\;}_{{\bf r}^\prime+{\bf r}}(\tau)\,
b^\dagger_{{\bf r}^\prime}(0) \}\, \rangle \;.
\label{G}
\end{equation}
Here $\langle...\rangle$ stands for the thermal expectation value,
${\cal T}$ is the time-ordering operator, $-\beta /2 \le\tau \le
\beta /2$, and $b^\dagger_{{\bf r}}(\tau)$ is the Bose particle creation
operator in the Matsubara representation. For $\tau < 0$ ($\tau > 0$)
one is computing the Green function for a vacancy (interstitial atom).
If, over large distances and/or long times, $n(r)$ and $G({\bm r},\tau )$
decay exponentially, then the state is non-superfluid, i.e., insulating.
Moreover, if $n(r)$ and  $G({\bm r},\tau )$ are temperature-independent up to
a certain length/time scale, then one can claim that ground state properties
on the corresponding scales are being explored (this is most readily
seen in the path-integral framework). Finally, by performing simulations
in the grand canonical ensemble one can investigate whether the solid
is commensurate or has gapless vacancies/interstitials in the ground state.

All of the above criteria have been used in recent first-principles
simulations of {\it hcp} helium crystals. They are based on the 
standard pairwise interatomic potential for helium \cite{aziz79} which 
essentially did not change in the last decades and is known to capture 
all important properties of condensed and gaseous helium with the relative 
accuracy of the order of one percent.
The study of exchange cycles\cite{Feynman}
reveals that they are extremely rare in the solid phase and indicative of
the insulating behavior \cite{bernu}. One may recall that macroscopic exchange
cycles are necessary for superfluidity, since
\[
n_s=\frac{mT\langle {\bm W}^2 \rangle} {dL}
\]
where $m$ is the helium mass, $d$ is the dimensionality, $L$ is the linear system size (periodic boundary conditions are assumed), and ${\bm W}=(W_x,W_y,W_z)$ are winding numbers, counting
how many times exchange cycles involving many particles wind around the system boundaries\cite{PC}. To simplify, we ignore here the tensor structure of $n_s$ in the crystal.
Initial concerns about ergodicity problems of simulations
probing large exchange cycles \cite{bernu} were eliminated in
subsequent path-integral Monte Carlo (PIMC) simulations based
on the Worm Algorithm \cite{superglass}, which established that the
superfluid fraction $n_s/n$ is unobservably small
(below $10^{-9}$), even in finite-size crystals comprising
only 800 atoms.

\begin{figure}[t]
\includegraphics [angle=-90, scale=0.22]{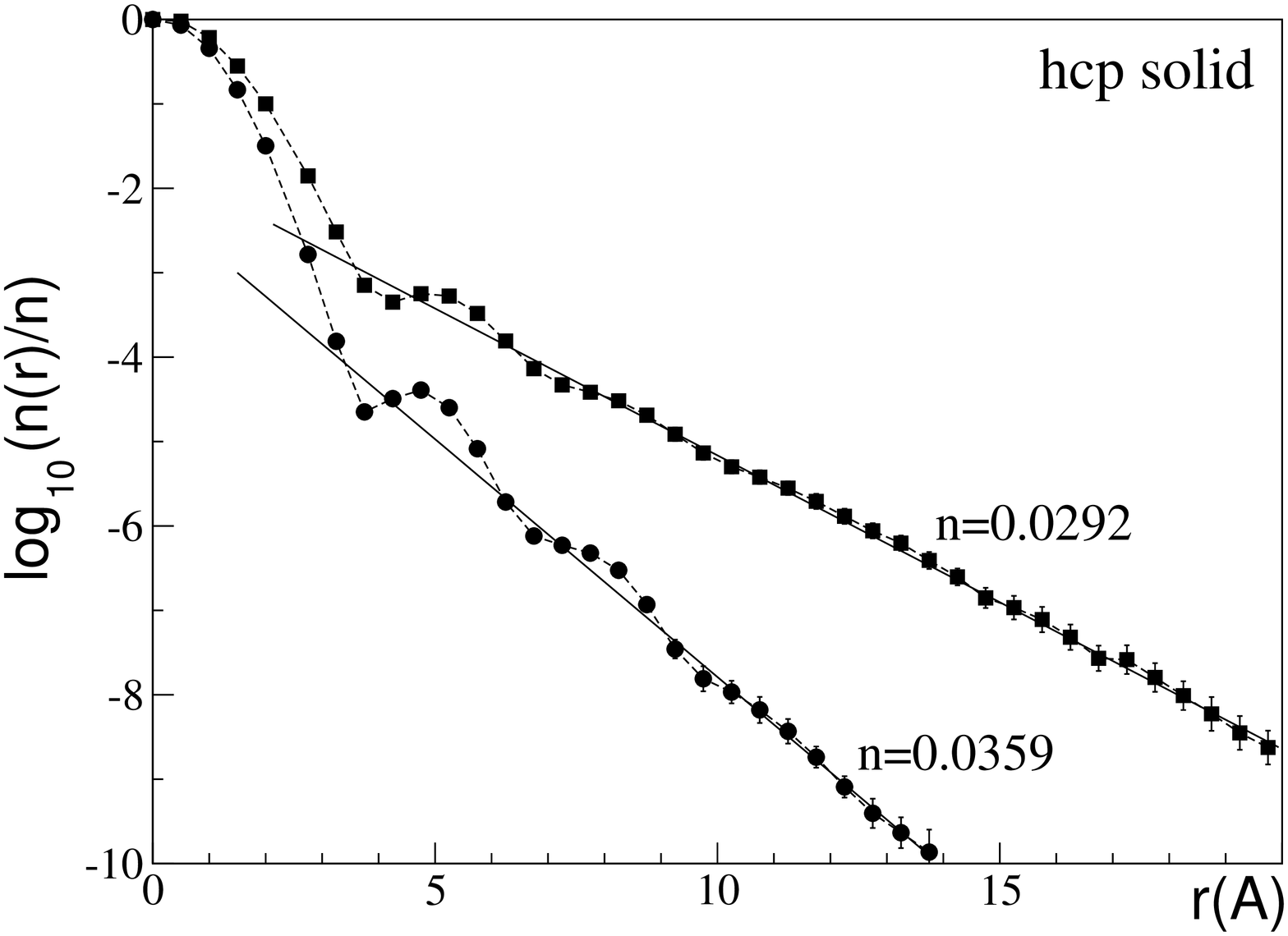}
\includegraphics [angle=-90, scale=0.230]{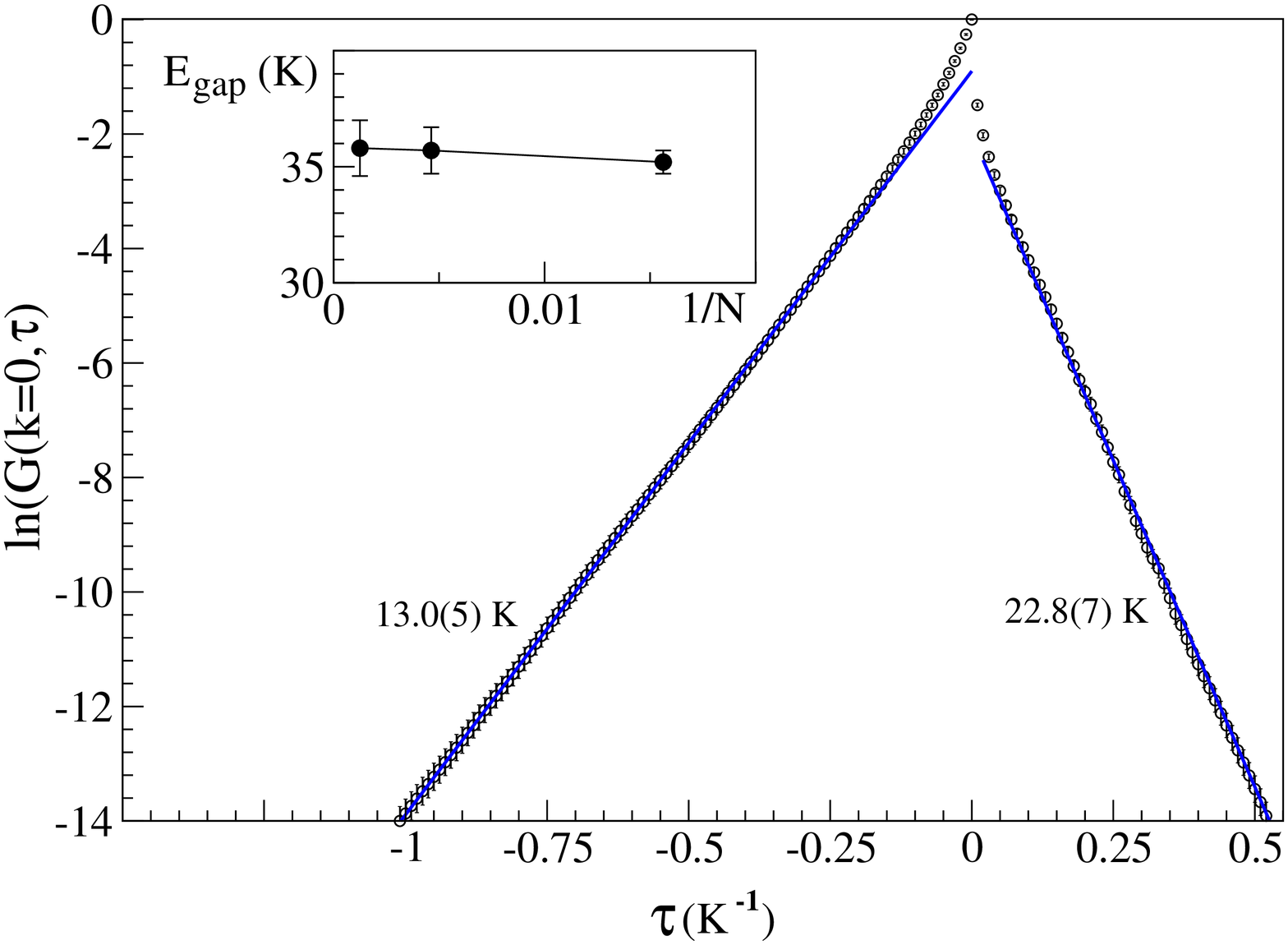}
\caption{Simulation results for the ideal hcp crystal of $N$=800 atoms
at low temperature $T=0.2$~K. Left panel \cite{superglass}:
the density matix close to the melting curve, $P=32$ bar ($n=0.0292$ \Am3 ),
and at elevated pressure $P=155$ bar ($n=0.0359$ \Am3 ).
The solid line is representing an exponential decay.
Right panel \cite{Pollet}: the zero-momentum Green function at 
the melting density $n_{\circ}$=0.0287 \Am3~.
Symbols refer to numerical data, solid lines are fits to the
long-time exponential decay.
The given numerical values are the interstitial ($\Delta_{\rm I}=22.8 \pm 0.7$ K)
and the vacancy ($\Delta_{\rm V}=13.0 \pm 0.5$ K) activation energies,
inferred from the slopes of $G$. The inset shows the vacancy-interstitial
gap $E_{\rm gap} = \Delta_{\rm I}+\Delta_{\rm V}$ for different system sizes.  }
\label{fig2}
\end{figure}

More detailed evidence for the insulating behavior of helium crystals
is provided by computations of the single-particle density matrix and 
the zero-momentum Matsubara Green function\cite{superglass,Pollet}. 
In Fig.~\ref{fig2}, we clearly see an exponential decay of the one-body 
density matrix $n(r)$ at large distances, both in the vicinity of the melting curve
(pressure $P=32$ bar) and at the pressure of $P=155$ bar.
The observed behavior is independent of temperature for $T < 1.5$ K, and there is
good agreement between independent calculations of $n(r)$
performed by different groups at the melting pressure \cite{clark}.

From the exponential decay of
the zero-momentum $G({\bf k}=0,\tau)$ (left panel of Fig.~\ref{fig2}),
one can obtain an activation energy
$\Delta =\Delta_{\rm V}\approx 13$ K for a vacancy
($\Delta_{\rm I}\approx 23$ K for an interstitial) by studying
\[ G({\bf k}=0,\tau)
\propto e^{-|\tau| \Delta}\;, \;\;\; |\tau | \to \infty ,\]
for  $\tau < 0$ ($\tau > 0$).
We note that these energies are quite large compared to the characteristic temperature
of $0.2$~K of the KC experiment, all but ruling out the possibility that
thermally excited vacancies may be behind the supersolid phenomenon
(it should also be noted that $\Delta_{\rm V}$ and $\Delta_{\rm I}$
increase with pressure) \cite{Pollet}.
This conclusion is consistent with the absence of
thermally activated vacancies at low $T$, as inferred from impurity 
mobility experiments (summarized in Ref.~\cite{fraass}), as well as
high-precision studies of the liquid-solid phase diagram \cite{Todosh}.
One may note that simulations of gaps in a system comprising 800 atoms are
essentially probing the thermodynamic limit, as finite-size effects
are negligible (see inset of left panel of Fig. \ref{fig2}).

The analysis of the X-ray data at low temperature in terms
of activated vacancies \cite{fraass} was questioned in Ref.~\cite{anderson}.
Specifically, the formula
\[n_v=e^{-\Delta_V/T}\ \biggl(\frac{mT}{2\pi}\biggr)^{3/2}\,,\]
with reported activation energies, in combination with the calculated
\cite{Galli_v} vacancy effective mass, can not account for the observed
change of the lattice constant (the effect would be too small).
However, since energies of optical and acoustic phonons at the
Brillouin zone boundary are also $\sim$ 10 K, one can not exclude
a conventional explanation in terms of crystal anharmonicity.
It seems, that all one can state with relative confidence is that the
vacancy concentration is below 1\%.

In the most conservative approach, the above-mentioned theoretical
results show that commensurate {\it hcp} crystals of \he4 are insulating and
locally stable since single vacancies have a finite energy cost.
This does not exclude, strictly speaking, a remote possibility that the
true ground state may have a finite, albeit small
concentration of vacancies\cite{Dai,anderson}.
Such scenarios, however, are ruled out by strong attractive interaction
between vacancies; indeed,
when vacancies are forcibly introduced in the crystal at any finite
concentration (in a ``computer experiment"), they immediately phase separate
from the crystal bulk \cite{Pollet}.
Under realistic experimental conditions, they will quickly
anneal at grain boundaries and dislocations; if the latter are absent,
then vacancies will form liquid droplets (at the melting curve) or
dislocation loops. In any case, the net result is that the commensurate
insulating {\it hcp} crystal is the true ground state.

In addition to the original picture of the dilute zero-point vacancy gas
\cite{Andreev69,Chester}, over the past few years there have been several theoretical proposals for the ideal supersolid phase of helium
\cite{tiwari,Baskaran,Dai,Hui,Ye}, differing in the microscopic
mechanisms of supersolidity. Some of the early works suggesting
commensurate SFS phases, were subsequently amended, to respect
the theorem that such phases should necessarily feature gapless
vacancies (not necessarily in the dilute gas phase).
In this respect, Refs.~\cite{tiwari,Hui} are still in error.
An interesting idea was proposed in Ref.~\cite{Dorsey}, that {\it hcp}
crystals may become SFS only under anisotropic stresses, which are
likely to occur in the experiment.
However, simulation results for {\it hcp} crystals at the melting curve,
show that vacancy and interstitial gaps hardly change, even under
very large anisotropic stress \cite{GBweb}. It appears at this point that
phenomenological scenarios and mean-field theories of supersolidity
do not work for \he4, and one has to seek the explanation of
recent experimental results outside the paradigm of the
homogeneous {\it hcp} crystal.

Before reviewing crucial experimental facts, let us
discuss the most important shortcomings (including some unphysical properties)
of Jastrow, as well as of the related ``shadow"\cite{SWF} variational
wave functions, in the context of helium.
Calculations based on these wave functions predict both (i)
a very small, but finite, condensate fraction for solid \he4 near the melting curve
(below $10^{-4}$) and (ii) large activation energies for vacancies
in solid helium samples comprising several hundred atoms\cite{galli}.
In the presence of short-range interactions among atoms, the physics
of vacancy formation is local in nature. All of the energy
contributions quickly approach their thermodynamic limit value,
as the system size is increased.
Indeed, the kinetic energy of localization in a volume $L^3$ is
proportional to $1/L^2$ (this contribution is not present in a system with periodic
boundary conditions). The direct coupling of the removed atom with
the rest of the system converges as $1/L^3$ for the Van der
Waals interaction; the same law describes convergence of
the deformation energy. The fast disappearance of finite-size effects
as $L$ is increased, is  observed in essentially all numerical
simulations of condensed helium (see, e.g. inset in Fig.~\ref{fig2}),
including those based on the variational approach.
On the other hand, vacancies are gapless in a state described by a
Jastrow wave function by construction!
It means that if variational calculations were taken to larger system
sizes they would have predicted negative corrections to the vacancy
energy which {\it increase}, not decrease, with $L$, making
$\Delta_I$ negative for $n_0L^3 >1$.
This unphysical behavior is only possible if irreducible
long-range multi-particle interactions, which can not be reduced to the
effective chemical potential shift, are present in the
system---something that is not observed in Nature.

The other problem with the Jastrow variational {\it ansatz}
(and, of other such wave functions as well)
is that it does not easily describe the physics of phase separation.
Because, as shown above, a dilute Bose gas of vacancies in helium
has attractive interactions and thus is an unstable, phase separating system,
any discussion of BEC and superfluidity of the uniform vacancy gas
under such circumstances is meaningless.

\section{$^4$He ``ice cream''}
 
In this Section, we shall argue that many experimental facts suggest that
the observed supersolid phenomenon in \he4 is induced by defects,
or disorder, in the torsion oscillator samples. This point of view
is supported by several numerical simulations, as well as by the recent 
direct observation of the grain boundary superflow \cite{Balibar}.

Let us assume that the observed drop in the resonant period of the
torsion oscillator can be attributed to a non-zero superfluid
response at a given frequency. This seems plausible, as the period
drop or, equivalently, mass decoupling increases as the oscillator
amplitude $A_{osc}$ and velocity $v_{osc}$ decrease. [To understand
why $n_s$ decreases with $v_{osc}$, one has to consider the
non-linear response of vortex loops and pinned vortex lines to the
flow.] In normal viscous media, an opposite correlation between the
mass decoupling and $v_{osc}$ is expected. One may also
attempt to understand the NCRI effect in terms of
kinetic relaxation or mass redistribution in the cell
\cite{ChanH2,Nussinov}. However, in helium samples, these explanations
face the crucial blocked-channel test \cite{ChanH2}---only
superflow can be sensitive to the sample topology on macroscopic
length-scales. It is also hard to believe that mass
redistribution kinetics in solid helium is reversible at low
temperature, i.e. the signal would not be reproducible upon heating
and cooling (assuming that $T<1~K$ in the cycle to avoid possible
annealing effects).

With the superfluid response in mind, we observe that many features
in the data simply do not fit into the homogeneous supersolid picture.
In particular:

\bigskip\noindent
\underline{\it Dependence of the superfluid density on temperature} \\
It is a well established fact that in a continuous normal-superfluid
(N-SF) transition, the superfluid density dependence on temperature
near the transition point at $T_{\rm c}$ is given by $(T-T_{\rm
c})^\nu$, where $\nu \approx 0.671(1)$ is the correlation length
exponent for the 3D XY-universality class \cite{nu}. So far, no
exception to this law, which predicts an {\it infinite} derivative
$dn_s/dT$ at $T_{\rm c}$ has been found, either theoretically or
experimentally, and the same is expected for the transition to the
SFS state \cite{Dorsey}. On the other hand, in all torsional
oscillator experiments displaying mass decoupling at low $T$, the
rise of $n_s(T)$  at low $T$ starts off with {\it zero} derivative
near $T_{\rm c}$, in turn rendering a precise determination of
$T_{\rm c}$ rather ambiguous. The most straightforward explanation
for such a gradual increase of the superfluid density would be a
broad distribution of local transition temperatures $T_{\rm c}({\bm
r})$ within a strongly inhomogeneous sample. The shape of the
$n_s(T)$ curve is then controlled by the probability distribution of
transition temperatures. Diffraction experiments might provide 
information on the crystal quality in experimental samples 
but these were not reported yet. 

\bigskip\noindent
\underline{\it Specific heat anomaly} \\
The other puzzling feature, which does not conform to the
established picture of the continuous
N-SF transition, is the absence of the specific heat maximum.
Given experimental uncertainties \cite{KC2}, the possible amplitude
of the specific heat anomaly is orders of magnitude too small for the
observed amount of the superfluid density \cite{balatsky2}.
Moreover, it was found that the dependence of the specific heat $C_v$ on
$T$ is linear in the supersolid regime. A linear $C_v(T)$
may originate from one-dimensional Luttinger-liquid type structures,
e.g., superfluid dislocation cores \cite{balatsky2} and
ridges between grain boundaries \cite{theorem,GBweb},
or just from a collection of two-level systems in the amorphous
sample \cite{philips}.

\bigskip\noindent
\underline{\it Dependence on geometry} \\
If the supersolid phenomenon observed in recent experiments is indeed a homogeneous
bulk effect, it is difficult to explain the observed dependence
of the microscopic\footnote{Microscopic is defined here as occurring on
length scales much larger than the correlation length but much smaller
than the system size} parameter $n_s$
on the geometry of the experimental setup . The scatter of  the reported
values of $n_s$ (spanning almost a decade), depending on annulus or open geometry
in the KC experiment, can be in principle explained if the sample quality,
and thus the distribution of $T_{\rm c}$, is strongly dependent on
the cell geometry (certainly a plausible argument).

\bigskip\noindent
\underline{\it Dependence on sample ``history''} \\
Rittner and Reppy \cite{Rittner} have reported the ``elimination of supersolid by annealing'',
in an experimental setup very close to that of Kim and Chan.
The other important observation, is that the torsional oscillator quality factor,
or inverse dissipation rate,
is higher in normal samples. This result should be regarded as
direct evidence in favor of disorder-induced supersolidity. However, such annealing
effects have not been confirmed by Kim and Chan \cite{KC3}, and the explanation for this
discrepancy is lacking, at the time of this writing.

\bigskip\noindent
\underline{\it \hee3 effect} \\
Bosonic superfluidity is a very robust phenomenon, and one does not expect
any significant changes in the superfluid properties when \hee3 impurities are added,
at a concentration $n_3 << n_s$. It is well established
\cite{Mullin,he3_1,Mikheev} that \hee3 substitution
atom in the \he4 solid matrix is described by a tight-binding
model, with a tiny hopping amplitude $J \sim 10^{-4}$ K. The small value of $J$
is consistent with the characteristic energy for nuclear magnetism in
solid \hee3, which is believed to be due to exchange of helium atoms\cite{Guyer}.
This result is important in several ways. First, it proves that \hee3 atoms
do {\it not} induce vacancy formation in their vicinity since the \hee3-vacancy
complex would be extremely mobile (the vacancy hopping amplitude
is nearly four orders of magnitude higher \cite{Galli_v}). Second, a small value of
$J$ makes \hee3 atoms extremely sensitive to virtually any type of
crystalline disorder, since defects will certainly create tight bound states
for \hee3. Even a weak deformation potential between \hee3 impurities is sufficient
to bind them \cite{Mullin,Andreev76}, or localize dilute \hee3 solutions
\cite{Kagan84}.

It is therefore very surprising, that both the
superfluid fraction and the onset of the superfluid response observed in KC experiments,
are {\it significantly} altered
when \hee3 impurities are added, even if the concentration $n_3$ is as small as few $ppm$,
(or even few $ppb$ !). The mismatch between $n_s$ and $n_3$ is so large that
it casts serious doubts on the homogeneous supersolid scenario \cite{theorem},
even on phenomenological grounds\cite{balatsky}.

In a disordered sample, the lighter \hee3 atom will try to minimize
its kinetic energy by binding to static defects which have lower
local particle number density, such as dislocation cores, grain
boundaries and ridges between them (vacancies are mobile and it is
not clear whether they can form a bound state with \hee3 atoms). One
thus expects an accumulation of \hee3 on defects at low temperature;
it is plausible that the local concentration of isotopic impurities
may be fairly high in certain regions of space. The other, somewhat
speculative, effect of \hee3 substitution might be that the quality
of samples, i.e. the amount of disorder, is itself a function of
$n_3$. For example, by concentrating at the perimeter of the
microcrystal,  \hee3 atoms may inhibit fast crystal growth, thereby
helping to make the ``ice cream''.

If we now couple these considerations with the theoretical
conclusion that only disorder can be responsible for the supersolid
phenomenon in helium, we find a very subtle interplay between the
superfluid response and \hee3 content. Imagine for a moment, that
superfluidity occurs along grain boundaries, while \hee3 is
concentrated at the boundary ridges. Assuming that \hee3 atoms
suppress superfluidity, we arrive at the picture where \hee3-rich
ridges form a Josephson junction network completely surrounding
grain boundaries. In this scenario it is conceivable that even a
tiny amount of \hee3 (per volume)  might be relevant for
supersolidity \cite{theorem}.

\bigskip\noindent
\underline{\it Dependence on pressure} \\
The obvious expectation is that supersolid properties
ought to be suppressed, as pressure is applied to the sample.
Applying pressure indeed drives down
the superfluid transition temperature $T_{\rm c}$.
Practically all known properties of helium crystals
are consistent with the overall tendency of helium atoms to behave more
classically at higher densities (see, e.g. Fig.~\ref{fig2}).
Surprisingly, the KC data \cite{KC3} show hardly {\it any} pressure dependence
for $n_s$ and $T_c$, up to $P\approx 65$~bar\footnote{It should be mentioned
that experimental data published so far are rather noisy}; at higher pressures,
the supersolid signal weakens, and apparently disappears at $P>160$~bar.
At this point one may wonder whether experimental data on
pressure dependence should be attributed to properties of
the {\it same} sample, as if the pressure was changed by squeezing
the volume, or different point represents properties of {\it different}
samples. If supersolidity is directly related to the sample quality,
which, in turn, depends on initial conditions and the solidification
protocol, then  data become very ambiguous, and their interpretation intricate.

While the above mentioned experimental facts do not fit into the homogeneous
supersolid crystal picture, there are others which are hard to reconcile
with {\it any} existing theoretical framework. For example:

\bigskip\noindent
\underline{\it ``Critical velocity''} \\
Superfluid decoupling in the torsion bob has the most unusual
dependence on the oscillator amplitude/velocity. Though in the
current setup velocity and amplitude are strictly related, it is
assumed that forces equilibrate in the sample fast enough to
attribute all changes in the superflow to the velocity $v$. It is
found that $n_s$ saturates at macroscopically small velocities,
$v_{\rm sat} \sim 10~\mu m/s \sim \hbar/mL$ (related to just few
circulation quanta). These values clearly have nothing to do with
the naive sound velocity estimate $c \sim \sqrt{n_sU/m} \sim 10~m/s$
based on typical helium parameters. The discrepancy is so large,
that there is virtually no room left for explaining $v_{\rm sat}$
using microscopic mechanisms. On the other hand, if extended defects
in the superfluid phase order are involved, e.g. vortex lines of
length $L$, then one faces the problem that the vortex motion
timescale $mL^2/\hbar^2$ is $5$ to $6$ orders of magnitude too long,
when compared to the oscillation period.

\bigskip\noindent
\underline{\it Zero-flow experiments} \\
Superfluidity is associated with an anomalous mass current
response to a gauge phase gradient or chemical potential difference.
There have been numerous attempts in the past to detect superflow in the solid
phase with negative results (see Ref. \cite{Meisel}). The same conclusion
was reached in more recent studies \cite{Beamish}, both in Vycor and
in bulk samples. As they stand now, flow experiments do {\it not} support,
and indeed are incompatible  with the supersolid phase of helium.
We can only note here that in disordered solids large pressure gradients are
common, and it is not obvious what chemical potential difference is applied
to the superfluid component in the disordered sample. For example, air pressure
in the cave deep under the mountain is still at a modest 1 bar value.

\subsection{Superglass}
Aside looking at defects that are typical for the polycrystalline sample,
one may wonder if helium can exist in an ``ultimate'' disordered solid
phase, namely, a glass. All known structural glasses are in a normal,
non-superfluid state.
In this regard, \he4 offers an intriguing possibility, unique for a
quantum solid, of being in a ``superfluid glassy" phase, or superglass (SG).
By definition, in the metastable SG phase the translation invariance
is broken, but the pair correlation function
$g({\bm r})=\langle n({\bm r})n(0)\rangle $
features no diagonal long range order. At the same time, a superglass
has ODLRO and non-zero superfluid response. Strange as it is, the SG
is reminiscent of a sponge soaked in a superfluid liquid made of the
same atoms.

Superglass was observed in numerical simulations of high-pressure
(about 150 bar) samples prepared by fast temperature quench from the
normal liquid state \cite{superglass}. Clearly, the dynamics of real
helium under cooling is vastly different from the thermalization
dynamics in the Monte Carlo simulation done (i) in imaginary time,
and (ii) in the absence of energy conservation. Thus, numerical
simulations rather answer the question of existence of the
metastable phase and its quasi-equilibrium properties, but not how
easily it can be prepared experimentally. One may also wonder if
helium SG is merely an artifact of the simulation algorithm (though
we do not see any obvious reason to suspect that), or a long-lived,
metastable, physical phase responsible for the unexpected outcome of
the acoustic-pulse experiment \cite{acoustic_pulse} which aimed at
nucleating the crystal phase in the middle of the cold, $50~mK$,
superfluid liquid but failed. The local pressure in the pulse
reached 160 bar, well above the theoretically predicted threshold of
60 bar for quantum nucleation \cite{Caupin}. It is not known what
the normal-superfluid transition temperature in the glassy phase is,
at what temperature the metastable liquid freezes into a glass, and
what role (if any) the SG might play in the torsion oscillator
experiments, e.g., by being trapped in some amount between
micro-crystals.

\section{Grain boundary superfluidity}

Even if the bulk phase of the material is an insulating solid,
there is no physical reason why grain boundaries (GB) in the same
material ought  to be insulating as well, especially in the vicinity
of the (weakly first-order) liquid-solid transition. It means that
SF in lower dimensionality can be obtained by simply placing two solid
pieces right next to each other. Conceptually, this possibility was
demonstrated in model simulations of domain walls in the checkerboard
solid formed by hard core bosons with the nearest-neighbor
repulsion on a simple cubic lattice \cite{SFwalls}. Subsequent
path-integral Monte Carlo simulations of \he4 polycrystals
\cite{GBweb} did reveal that some (not all !) grain boundaries
are likely to support superfluidity. Superfluidity in the
layer at the disordered Vycor substrate was also reported in
Ref.~\cite{Khairallah}.

\begin{figure}[t]
\includegraphics [angle=-90, scale=0.26]{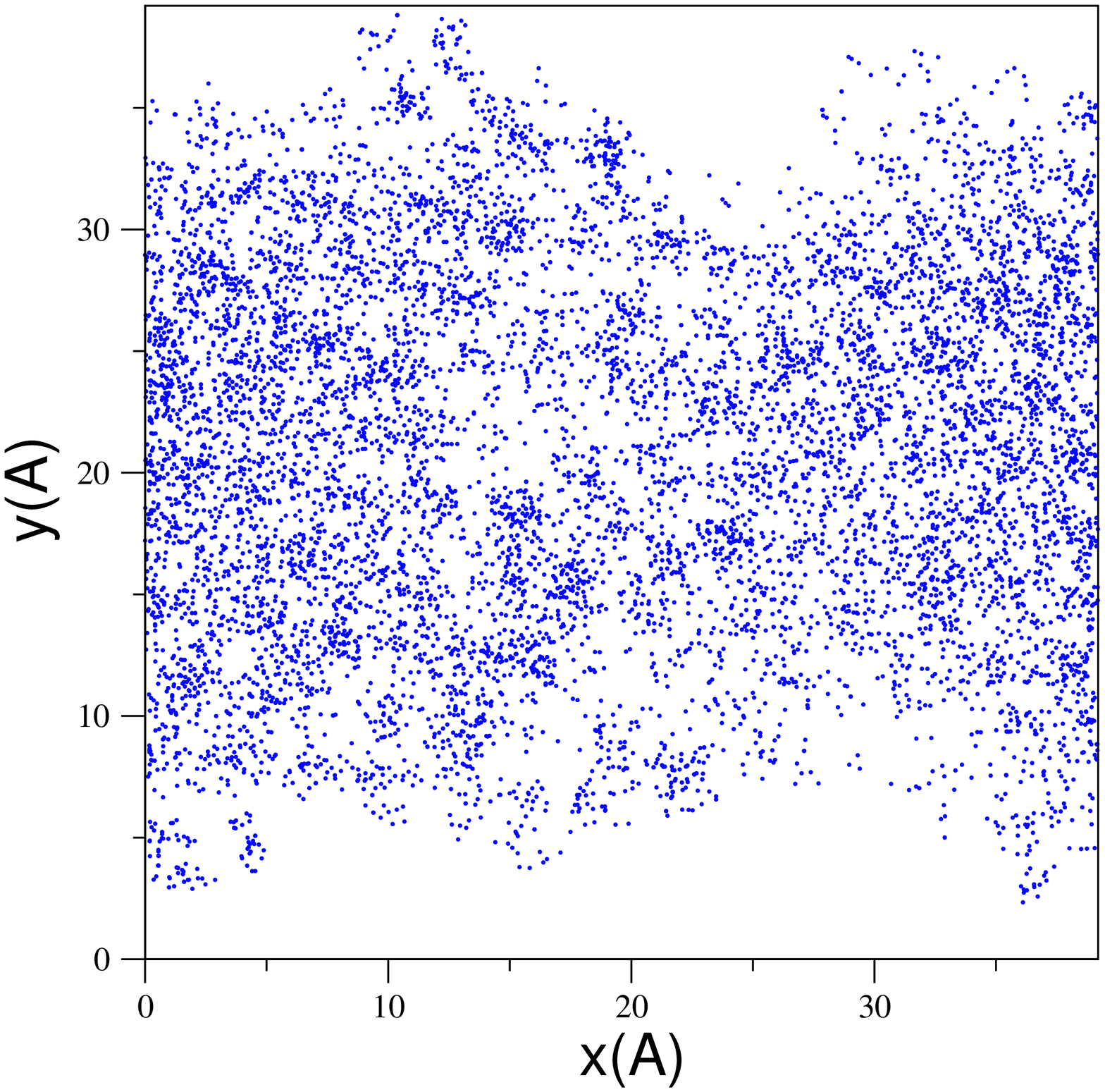}
\includegraphics [angle=-90, scale=0.26]{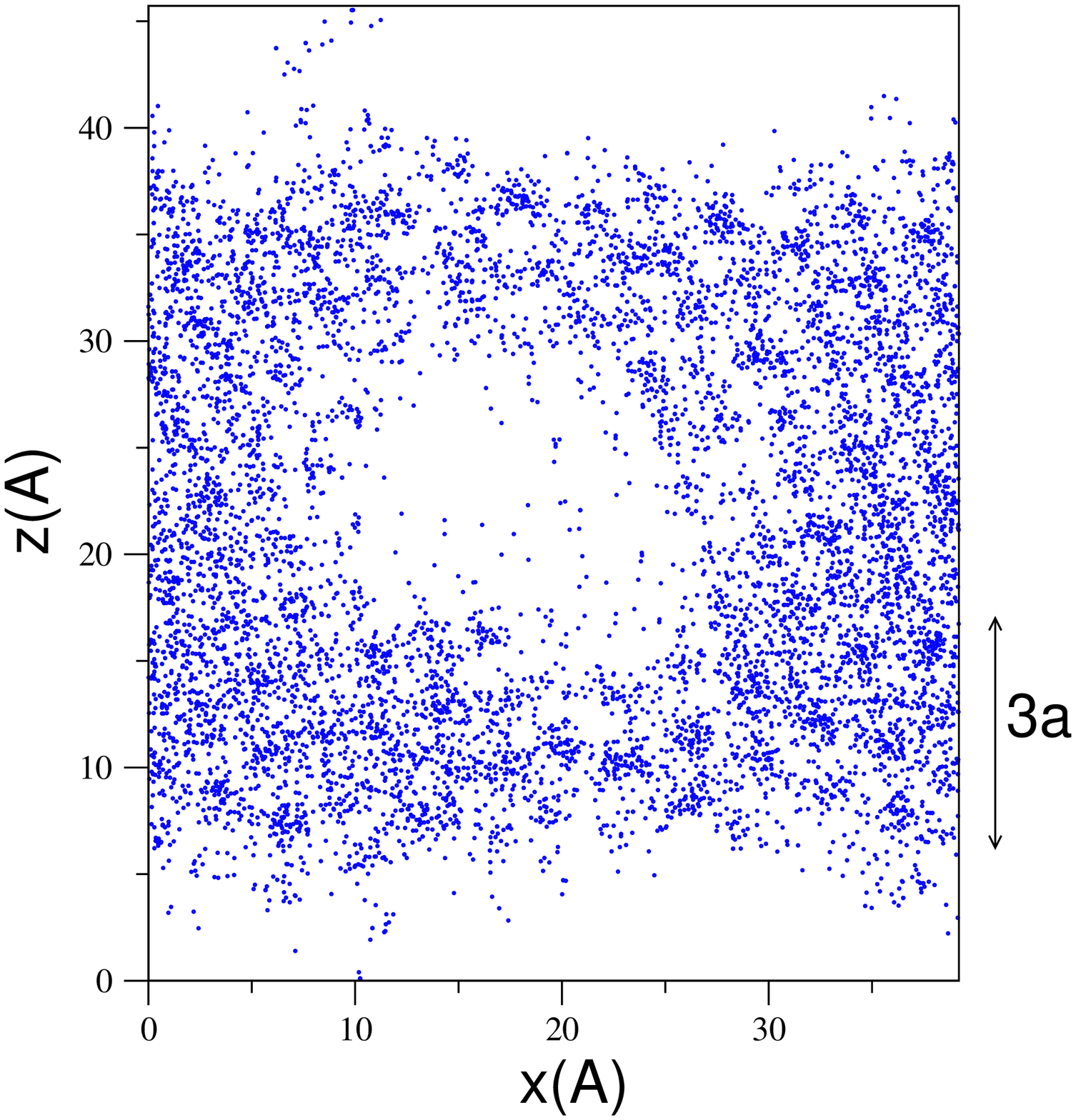}
\caption{Phase coherence properties of grain boundaries and ridges in the
\he4 sample consisting of two microcrystals with about $12\times 12 \times 7$
atoms each. Periodic boundary conditions are assumed.
The XY boundaries between different crystals were created by rotating
both of them at random. Other grain boundaries are induced by periodic
boundary conditions in the simulation box of rectangular shape, i.e.
they are created within the same crystal. We forced YZ boundaries to be
insulating by fixing atom positions in the YZ-layer ($1.5~a$ thick)
at the equilibrium hcp points. Superfludity was then possible only
along the $\hat{x}$ and $\hat{z}$ directions.
Points in the figure are show positions of particles which participated in
macroscopic exchange cycles with non-zero winding numbers, i.e. directly
contributed to the superfluid response. The two panels (from left to right)
projections on the XY (only the upper half of the sample is projected on
the XY plane) and XZ planes.
The Monte Carlo simulation was performed
in the grand canonical ensemble at the coexistence curve.}
\label{fig3}
\end{figure}

Simulations also find that nearly all ridges (lines of contact
between different GB) show robust phase coherence properties. It is
thus possible that superfluidity across grain boundaries of small
(216-atom) crystals is nothing but the proximity effect. This issue
was addressed in simulations of much larger GB \cite{GBweb} in a
system of about 2000 atoms consisting of two randomly oriented
crystallites at the melting density $n=0.0287$~\Am3 . The main
conclusions did not change; while grain boundaries at special angles
and high symmetry directions are insulating, generic grain
boundaries appear to be superfluid with typical transition
temperatures (orientation dependent) of about half a Kelvin and the
maximum possible $T_c$ at about $1.5$ K. The width of the superfluid
GB region is about $\sim 3a$ where $a$ is the nearest-neighbor
distance in the hcp crystal, see Fig.~\ref{fig3}.

In a remarkable recent experiment \cite{Balibar}, superflow along the grain
boundary was detected in an experiment whose setup is similar to that shown in
Fig.~\ref{fig1}. The difference between the solid levels was found to decay
according to the linear law $\dot{h}=const$ characteristic of the superfluid
flow at the critical velocity (estimated to be of the order of meters per second).
The flow was detected only in the presence of grain boundaries; otherwise
the sample did not relax. On one occasion, the dynamics abruptly came to a halt,
as the boundary had suddenly disappeared. A grain boundary is a topological
crystalline defect, and its quick (in less than $1$ min time) disappearance
is by itself an amazing observation, since it has no obvious explanation
other than suggesting that boundaries are in a rough
superfluid phase, supporting coherent recrystallization waves.

\section{Perspectives for future theoretical and experimental work}

It is currently our strong belief, that all known experimental facts,
together with
results of first-principles numerical simulations, rule out any explanation
of the observed supersolid phenomenon in \he4 within a homogeneous
crystal framework. The only reasonable alternative, consistent
with many observations, is that of supersolidity induced
by crystalline disorder. At the moment, very little is known for sure
about structure properties of solid samples in the torsion oscillator
experiments. Hopefully, in future investigations it will be possible to
control sample quality {\it in situ}, e.g. by forming solids under constant
pressure, rather than under constant volume conditions.
Helium crystals of high quality are usually made out of the superfluid liquid
at constant pressure. Solids grown from normal liquid under constant volume
conditions are not transparent (similar in appearance to milk) \cite{Bob}.
It is also desirable to have direct information on the solid order
in the same supersolid sample, e.g., by optical means.
Even more urgently, several experimental observations of different
groups have to be reconciled, especially regarding
pressure dependence and annealing effects.
It seems also that additional measurements of the specific heat
and other thermodynamic properties can help in quantifying
the amount of ``disorder'' in experimental samples \cite{balatsky2}.

Analytic theories of the SFS state in continuous space are phenomenological 
in nature when it comes to predicting, for a particular Hamiltonian,
whether this state exists and what are crucial parameters for the SFS state 
in terms of bare couplings. Supersolids arise from competition between the kinetic 
and potential energy terms invalidating perturbative and mean-field treatments.  
At present, we do not know a
single example of a system characterized by a realistic interatomic potential,
which has a stable SFS phase in continuous space.
This work is worth pursuing, since there
are exciting new possibilities in engineering interparticle
interactions in cold atomic systems. It would be also interesting
to see, once the ideal SFS phase is found, which of the recent
theoretical predictions capture the microscopic picture
of the supersolid phase more adequately.

\begin{figure}[t]
\includegraphics [angle=-90, scale=0.4]{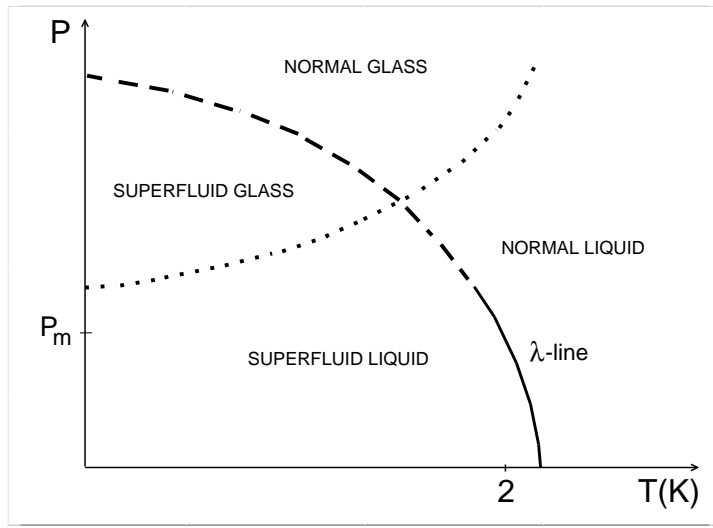}
\vspace*{-1cm}
\caption{A sketch of the metastable helium glass phase
diagram. Apart from the melting pressure $P_m$ we do not show the
equilibrium liquid-solid line in the figure; the metastable glass
state is supposed to be inside the stable solid phase. Since the
liquid-solid transition is first-order the normal-superfluid
transition in the metastable liquid is a continuation of the $\lambda$-line.
The dashed line is indicating a sharp crossover, not the genuine
phase transition; above this line the structural relaxation
time is beyond the experimental reach.}
\label{fig4}
\end{figure}

So far, the only observation of the superfluid glass phase was
numerical (at $T=0.2~K$ and $n=0.0359$\Am3). Most properties of the
helium glass phase, and even its very existence, are a mystery.
Theoretically, more work has to be done to quantify the SG state,
and to see whether the heuristic phase diagram shown in
Fig.~\ref{fig4} takes place. Experimentally, one may think of
preparing glassy solid samples by fast density increase at low
temperature rather then temperature quench at high density.
Experimentally, it seems that in helium pressure increase can be
achieved on a much shorter time scale than cooling down to
$T=0.2~K$. [One objection to this proposal is that crystal growth
from the cell wall at low temperature is extremely fast
\cite{balibar3}.] Another way of looking for non-ergodic properties
of helium samples, i.e. system dependence on the intial conditions
and preparation protocol, is to compare samples cooled while
rotating with samples rotated after cooling \cite{Kubota2}.

At the moment, the grain boundary flow experiments give us the most
unambiguous clue to the solution of the supersolid puzzle. On the other hand,
a whole plethora of new questions arise in this connection.
The problem is very broad, not only because crystalline interfaces
can be prepared at different angles with respect to the
crystal axes, but also because one can get different results,
depending on how well the initial state was ``annealed'' (with
respect to minimizing the classical deformation energy), and whether
defects are ``doped'' with out-of-equilibrium vacancies and/or
interstitials (one may wonder if vacancies cluster on defects the same
way they do it in the bulk and whether they can make, say a dislocation core,
superfluid ?).

In the experimental setup of Ref. \cite{Balibar},
the superflow in the boundary layer was observed
only at the coexistence curve. It is important to extend this type
of studies to higher pressures, though the experimental design has
to be modified. One possibility \cite{sandwich}, is to
look at the same topology as in Fig.~\ref{fig1} with the bulk solid being
sandwiched between two Vycor plates. Since helium in Vycor
remains  superfluid at pressures about $10~bar$
higher than the bulk melting value $P_m$ one may think of
measuring the flow through the solid in response to the pressure
difference between the two arms of the U-tube.

Another fascinating subject is recrystallization waves in the
boundary layer. Though possible in principle \cite{SFwalls}
and being a logical explanation for the fast disappearance
of grain boundaries in the experiment \cite{Balibar}, recrystallization
waves still remain to be seen and detected. Formally, there is no
one-to-one connection between recrystallization waves and
superfluidity in the boundary, e.g. smooth domain walls
in the 3D checkerboard solid remain superfluid at zero-temperature,
but the theory of the interplay between the two phenomena
is basically non-existent.

Previous NMR studies of \hee3 solid solutions
\cite{Mullin,he3_1,Mikheev,Kagan84} provided detailed
information on the isotopic impurity hopping dynamics, as well as
on interparticle interactions. These studies were performed at relatively
high temperatures $T>0.7$~K and in high quality crystals.
It would be interesting to see if NMR techniques can be used to understand
what happens to \hee3 atoms in disordered samples grown from the normal
high-density liquid and cooled down to the $100$~mK range. If \hee3 atoms
form a dense solution at the defects, one should be able to see it in the
increased NMR relaxation rates since coupling between the \hee3 magnetic moments
is the only available mechanism. Another exciting possibility is to
use NMR techniques to see if \hee3 solids can be in the metastable
glass phase.

The anomalously low critical velocity in the torsional oscillator
studies is yet to be understood, even at the expense of admitting
that the observed phase is, in fact, not superfluid \cite{anderson}.

\section{Concluding remarks}

Supersolids may happen in two ways: either as ideal incommensurate
crystals with zero-point vacancies or commensurate crystals full of
topological defects.
Though the original observation of supersolidity in \he4 can not be
interpreted in terms of zero-point vacancy mechanism in a perfect
crystal, things turn out to be far more interesting and exciting
than expected. Instead of one well-characterized crystal, we now
face a whole variety of disordered solid samples. This immediately
brings into focus grain boundaries, dislocations, and glassy phases
which are, on their own right, new superfluid physical systems. For
example, dislocations and grain boundaries should be considered as
special low-dimensional objects inseparable from their
three-dimensional environment. Studies of their fascinating
properties make the experimental and theoretical landscape
multi-dimensional, since in addition to pressure and temperature,
the set of important parameters also includes various orientation
angles. There is little doubt that this research field is going to
last. We also believe that superfluid amorphous helium is a
unique and remarkable state which will add a new angle to studies
of glassy systems. Apart from understanding the role of crystalline
defects in the supersolid phenomenon, it is extremely important to
formulate what type of realistic interaction potential between
particles can lead to the ideal supersolid phase---this long-term
goal which started more than 50 years ago is yet to be achieved.
Existing experimental and theoretical tools are more than capable of making
significant progress in achieving these goals, both at the
phenomenological and microscopic levels.

\section{Acknowledgment}

I am indebted to B. Svistunov, M. Boninsegni and M. Troyer
for numerous discussions and valuable suggestions which lead
to this work. I acknowledge financial support by the National
Science Foundation under Grant No. PHY-0426881.


\begin{thebibliography}{99}

\bibitem{KC1} E. Kim and M. H. W. Chan,
Nature {\bf 427}, 225-227 (2004).
\bibitem{KC1b} E. Kim and M. H. W. Chan,
Science {\bf 305}, 1941-1944 (2004).

\bibitem{KC2} A.C. Clark and M. H. W. Chan,
J. Low Temp. Phys. {\bf 138}, 853-858 (2005).

\bibitem{KC3} E. Kim and M. H. W. Chan,
Phys. Rev. Lett. {\bf 97}, 115302 (2006).

\bibitem{Rittner} A. S. Rittner and J. D. Reppy,
cond-mat/0604528.

\bibitem{Shirahama} M. Kondo, S. Takada, Y. Shibayama, and K. Shirahama,
cond-mat/0607032.

\bibitem{Kubota} A. Penzyev, and M. Kubota,
unpublished, preliminary results are available online at
http://online.kitp.ucsb.edu/online/smatter\underline{$\;\;$}m06

\bibitem{Leggett70} A. J. Leggett,
Phys. Rev. Lett. {\bf 25}, 1543 (1970).

\bibitem{Penrose51} O. Penrose,
Phil. Mag. {\bf 42}, 1373 (1951).

\bibitem{Penrose} O. Penrose and L. Onsager,
Phys. Rev. {\bf 104}, 576 (1956).

\bibitem{Yang} C. N. Yang,
Rev. Mod. Phys. {\bf 34}, 694 (1962).

\bibitem{Leggett04} A. J. Leggett,
Science {\bf 305}, 1921 (2004).

\bibitem{Andreev69} A. F. Andreev and I. M. Lifshitz,
Sov. Phys. JETP {\bf 29}, 1107 (1969);

\bibitem{Chester} G. V. Chester,
Phys. Rev. A {\bf 2}, 256 (1970).

\bibitem{yes_SFS1} P. Sengupta, L.P. Pryadko, F. Alet,
                   M. Troyer, and G. Schmidal,
Phys. Rev. Lett. {\bf 94}, 207202 (2005);
G.G. Batrouni, F. Hebert, and R.T. Scalettar,
Phys. Rev. Lett. {\bf 97}, 087209 (2006).

\bibitem{yes_SFS2} S. Wessel  and M. Troyer,
Phys. Rev. Lett. {\bf 95}, 127205 (2005);
D. Heidarian and K. Damle,
Phys. Rev. Lett. {\bf 95}, 127206 (2005);
 R. G. Melko, A. Paramekanti, A. A. Burkov, A. Vishwanath,
D. N. Sheng, and L. Balents,
Phys. Rev. Lett. {\bf 95}, 127207 (2005);
M. Boninsegni  and N. Prokof'ev,
Phys. Rev. Lett. {\bf 95}, 237204 (2005).

\bibitem{Meisel} M. W. Meisel,
Physica {\bf B}: Cond. Matt. {\bf 178}, 121 (1992).

\bibitem{Beamish} J. Day, T. Herman, and J. Beamish,
Phys. Rev. Lett. {\bf 95}, 035301 (2005);
             J. Day and J. Beamish,
Phys. Rev. Lett. {\bf 96}, 105304 (2006).

\bibitem{Balibar} S. Sasaki, R. Ishiguro, F. Caupin, H. J. Maris, and S. Balibar,
 Science {\bf 313}, 1098 (2006).

\bibitem{Kohn} W. Kohn, Phys. Rev. {\bf 133}, A171 (1964).

\bibitem{Sherrington70} W. Kohn and D. Sherrington,
Rev. Mod. Phys. {\bf 42}, 1 (1970).

\bibitem{BKT}  V. L. Berezinskii, Sov. Phys. JETP {\bf 32}, 493 (1970);
{\it ibid } {\bf 34}, 610 (1971);
J. M. Kosterlitz and D. J. Thouless, J. Phys. C {\bf 6}, 1181 (1973);
J. M. Kosterlitz, J. Phys. C {\bf 7}, 1046 (1974).

\bibitem{Glyde} H. R. Glyde, R. T. Azuah and W. G. Stirling,
                Phys. Rev. B {\bf 62}, 14337  (2000).

\bibitem{ceperley95} D. M. Ceperley, Rev. Mod. Phys. {\bf 67}, 279 (1995).

\bibitem{moroni04}
S. Moroni and M. Boninsegni, J. Low Temp. Phys. {\bf 136}, 129 (2004).

\bibitem{theorem} N. Prokof'ev and B. Svistunov,
Phys. Rev. Lett. {\bf 94}, 155302 (2005).

\bibitem{Reatto} L. Reatto
Phys. Rev. {\bf 183}, 334 (1969)

\bibitem{aziz79}
R. A. Aziz, V. P. S. Nain, J. S. Carley, W. L. Taylor and G. T.
McConville, J. Chem. Phys. {\bf 70}, 4330 (1979).

\bibitem{Feynman} R. P. Feynman, Phys. Rev. {\bf 90} 1116 (1953); {\it ibid}
{\bf 91} 1291 (1953).

\bibitem{bernu} D. M. Ceperley and B. Bernu,
Phys. Rev. Lett. {\bf 93}, 155303 (2004).

\bibitem{PC} E. L. Pollock and D. M. Ceperley,
Phys. Rev. B {\bf 36}, 8343 (1987).

\bibitem{superglass}
M. Boninsegni, N. V. Prokof'ev and B. V. Svistunov,
Phys. Rev. Lett. {\bf 96}, 105301 (2006).

\bibitem{clark} B. K. Clark and D. M. Ceperley,
Phys. Rev. Lett. {\bf 96}, 105302 (2006).

\bibitem{fraass}
B. A. Fraass, P. R. Granfors and R. O. Simmons, Phys.
Rev. B {\bf 39}, 124 (1989).

\bibitem{Todosh} I. A. Todoshchenko, H. Alles, H. J. Junes,
                 A. Ya. Parshin, V. Tsepelin, cond-mat/0607081.

\bibitem{Galli_v} D. E. Galli and L. Reatto,
 Phys. Rev. Lett. {\bf 90}, 175301 (2003).

\bibitem{phonons} F. P. Lipschultz, V. J. Minkiewicz, T. A. Kitchens,
G. Shirane, and R. Nathans,
Phys. Rev. Lett. {\bf 19}, 1307 (1967)

\bibitem{Pollet} M. Boninsegni, A. Kuklov, L. Pollet,
                 N. Prokof'ev, B. Svistunov, and M. Troyer,
Phys. Rev. Lett. {\bf 97}, 080401 (2006)

\bibitem{Dai} X. Dai, M. Ma, and F.-C. Zhang,
Phys. Rev. B {\bf 72}, 132504 (2005).

\bibitem{anderson}
P. W. Anderson, W. F. Brinkman and D. A. Huse,
Science {\bf 310}, 1164 (2005).

\bibitem{tiwari}  M. Tiwari and A. Datta, cond-mat/0406124.

\bibitem{Baskaran} G. Baskaran, cond-mat/0505160.

\bibitem{Hui} Hui Zhai and Yong-Shi Wu,
J. Stat. Mech.  P07003 (2005).

\bibitem{Ye} J. Ye,
Phys. Rev. Lett. {\bf 97}, 125302 (2006).

\bibitem{Dorsey} A.T. Dorsey, P.M. Goldbart, and J. Toner,
Phys. Rev. Lett. \textbf{96}, 055301 (2006).

\bibitem{GBweb} L. Pollet, M. Boninsegni, A. Kuklov,
N. Prokof'ev, B. Svistunov, and M. Troyer, in preparation.
Preliminary results on superfluidity
of grain boundaries in \he4 were reported at the
KITP Miniprogram: {\it The Supersolid State of Matter},
February 6-17, (2006);
[$http://online.kitp.ucsb.edu/online/smatter_m06/svistunov$].

\bibitem{SWF} S. Vitiello, K. Runge, and M. H. Kalos,
Phys. Rev. Lett. {\bf 60}, 1970 (1988).

\bibitem{galli} D.E. Galli, M. Rossi, and L. Reatto,
Phys. Rev. B {\bf 71}, 140506(R) (2005).

\bibitem{ChanH2}  A.C. Clark, X. Lin, M.H.W. Chan,
                  cond-mat/0610240.

\bibitem{Nussinov} Z. Nussinov, A.V. Balatsky, M.J. Graf, S.A. Trugman,
                   cond-mat/0610743.

\bibitem{nu} J. A. Lipa, J. A. Nissen, D. A. Stricker, D. R. Swanson,
and T. C. P. Chui, Phys. Rev. B {\bf 68}, 174518 (2003).

\bibitem{balatsky2} A.V. Balatsky, M.J. Graf, Z. Nussinov, and S.A. Trugman,
cond-mat/0606203.

\bibitem{philips} W. A. Phillips,
J. Low Temp. Phys. {\bf 7}, 351 (1972);
P. W. Anderson, B. I. Halperin, and C. M. Varma,
Phil. Mag. {\bf 25}, 1 (1972).

\bibitem{Mullin} M.G. Richards, J.H. Smith, P.S. Tofts, and W.J. Mullin,
Phys. Rev. Lett. {\bf 34}, 1545 (1975).

\bibitem{he3_1} M. G. Richards, J. Pope, P. S. Tofts, and J. H. Smith,
J. Low. Temp. Phys. {\bf 24}, 1 (1976);
A. R. Allen, M. G. Richards, J. Schratter,
J. Low. Temp. Phys. {\bf 47}, 289 (1982).

\bibitem{Mikheev} V. A. Mikheev, V. A. Maydanov, and N. P. Mikhin,
Solid State Comm. {\bf 48}, 361 (1983).

\bibitem{Guyer} R.A. Guyer, R.C. Richardson, and L.I. Zane,
Rev. Mod. Phys. {\bf 43}, 532-600 (1971).

\bibitem{Andreev76} A. F.  Andreev,
Sov.Phys. - JETP {\bf 41}, 1170 (1976).

\bibitem{Kagan84} Yu. Kagan and L.A. Maksimov,
Sov. Phys. - JETP {\bf 60}, 201 (1984).

\bibitem{balatsky} A. V. Balatsky and E. Abrahams,
cond-mat/060253.

\bibitem{acoustic_pulse} F. Werner, G. Beaume, A. Hobeika, S. Nascimbene,
                  C. Herrmann, F. Caupin and S. Balibar,
J. Low Temp. Phys. {\bf 136}, 93 (2004).

\bibitem{Caupin}
F. Caupin, S. Balibar, and H. J. Maris,
Physica B {\bf 329}---{\bf 333}, 356 (2003).

\bibitem{SFwalls} E. Burovski, E. Kozik, A. Kuklov, N. Prokof'ev, and B. Svistunov,
Phys. Rev. Lett. {\bf 94}, 165301 (2005).

\bibitem{Khairallah} S.A. Khairallah and D.M. Ceperley,
Phys. Rev. Lett. {\bf 95}, 185301 (2005).

\bibitem{Bob} R. Hallock, private communication.

\bibitem{balibar3} S. Balibar, T. Mizusaki, and Y. Sasaki,
J. Low Temp. Phys. {\bf 120}, 293 (2000);
X. Chavanne, S. Balibar, and F. Caupin,
Phys. Rev. Lett. {\bf 86}, 5506 (2001);
J. Low Temp. Phys. {\bf 125} 155 (2001).

\bibitem{Kubota2} M. Kubota, private communication.

\bibitem{sandwich} R. Hallock and M. Ray,
in preparation.

\bibitem{Kolya}
H. P. B\"uchler, E. Demler, M. Lukin, A. Micheli, N. Prokofiev, G. Pupillo, and P. Zoller,
{\it Strongly Correlated 2D Quantum Phases with Cold Polar Molecules:
Controlling the Shape of the Interaction Potential},
cond-mat/0607294.

\bibitem{schneider} T. Schneider and C.P. Enz,
{\it Theory of the Superfluid-Solid Transition of \he4},
Phys. Rev. Lett. {\bf 27}, 1186-1188 (1971).


\bibitem{dipolar_OL} K. Goral, L. Santos, and M. Lewenstein,{\it Quantum Phases of Dipolar Bosons in Optical Lattices}, Phys. Rev.
Lett. 88, 170406 (2002).

\end{thebibliography}
\end{document}